\documentclass[twocolumn,aps,amsmath,amssymb,prb]{revtex4}
\usepackage{graphicx}

\renewcommand{\L}{{\cal L}}
\newcommand{\U}{{\cal U}}
\newcommand{\A}{{\cal A}}
\newcommand{\B}{{\cal B}}
\newcommand{\G}{{\cal G}}

\newcommand{\Sec}[1]{Sec.\,\ref{#1}}
\newcommand{\ti}{\tilde}

\newcommand{\be}{\begin{equation}}
\newcommand{\ee}{\end{equation}}
\newcommand{\bea}{\begin{eqnarray}}
\newcommand{\eea}{\end{eqnarray}}
\newcommand{\bsube}{\begin{subequations}}
\newcommand{\esube}{\end{subequations}}
\newcommand{\Eq}[1]{eq\,\ref{#1}}
\newcommand{\Eqs}[1]{eqs\,\ref{#1}}
\newcommand{\Fig}[1]{Fig.\,\ref{#1}}

\newcommand{\la}{\langle}
\newcommand{\ra}{\rangle}
\newcommand{\dd}{\ra\ra\la\la}
\newcommand{\kT}{\mbox{$k_{\rm B}T$}}

\begin{document}

\title{Kinetics and thermodynamics of electron transfer
  in Debye solvents:\\
   An analytical and nonperturbative reduced density matrix theory
}
\author{Ping Han,$^{a,c)}$ Rui-Xue Xu,$^{b,c)\ast}$
      Baiqing Li,$^{b,c)}$ Jian Xu,$^{b)}$
        Ping Cui,$^{b,c)}$
          Yan Mo,$^{c)}$
 }
\author{YiJing Yan$^{a,b,c)}$}
\email{rxxu@ustc.edu.cn; yyan@ust.hk}
\affiliation{$^{a)}$State Key Laboratory of Molecular Reaction Dynamics,
  Dalian Institute of Chemical Physics,
  Chinese Academy of Sciences, Dalian 116023, China \\
$^{b)}$Hefei National Laboratory for Physical Sciences
  at the Microscale, University of Science  and Technology of China,
  Hefei 230026, China    \\
$^{c)}$Department of Chemistry,
    Hong Kong University of Science and Technology, Kowloon, Hong Kong}
\date{\today}

\begin{abstract}
 A nonperturbative electron transfer rate theory
is developed based on the reduced density matrix dynamics, which
can be evaluated readily for the Debye solvent model without
further approximation. Not only does it recover for reaction rates
the celebrated Marcus' inversion and Kramers' turnover behaviors,
the present theory also predicts for reaction thermodynamics,
 such as equilibrium Gibbs free-energy and entropy, some
interesting solvent-dependent features that are calling for
experimental verification. Moreover, a continued fraction Green's
function formalism is also constructed, which can be used together
with Dyson equation technique, for efficient evaluation of
nonperturbative reduced density matrix dynamics.
\end{abstract}

\maketitle

\section{Introduction}
\label{thintro}
   Solvent environment plays a crucial role in
determining chemical kinetic properties.
Its interaction induces fluctuation that compensates
for the activation energy, and also results in relaxation
that stabilizes the reaction product.
This dual role of solvent interaction,
related via fluctuation-dissipation relation,
has been recognized since Einstein in
his study of Brownian motion.\cite{Ein05549}
 The effect of solvent interaction
on chemical kinetics was first studied by Kramers in his classical
Fokker-Planck-equation (FPE) approach to the rate theory of
isomerization reaction.\cite{Kra40284} This is a classical
 reduced equation-of-motion (EOM) approach,
 in which chemical reaction is
 treated as escape over barrier of particle moving in
 one-dimensional potential surface and subject
 to a Langevin force
 originated from stochastic solvent interaction.
 The resulting rate, as function of solvent viscosity,
 exhibits a turnover behavior:
 it increases linearly as viscosity initially,
 reaches a maximum at an intermediate viscosity value,
 and falls off inversely as viscosity in the high
 friction regime.\cite{Kra40284,Han90251}

    Electron transfer (ET) processes
constitute another class of systems whose dependence on solvent
environment has been extensively studied since Marcus' pioneering
contribution.\cite{Mar56966,Mar64155,Mar85265}
 Consider the simplest ET system in which
the total Hamiltonian reads
 \be \label{HT}
 H_{\rm T} = h_a |a\ra\la a| + (h_b+E^{\circ})|b\ra\la b|
     + V(|a\ra\la b|+|b\ra\la a|).
\ee Here, $h_a$ and $h_b$ are the solvent Hamiltonians for the ET
system in the donor and acceptor states, respectively,
$E^{\circ}$ is the reaction endothermicity, and $V$ the transfer
coupling matrix element. The system is initially in the donor
$|a\ra$ site, with the solvent (bath) equilibrium density matrix
$\rho_a^{\rm eq}\propto e^{-h_a/(k_{\rm B}T)}$ at the temperature
$T$. The reaction coordinate is now $U\equiv h_b - h_a$,
 which is purely of solvent in nature
 and called the solvation coordinate.
  Its static mean,
   $\lambda=\la U\ra \equiv {\rm tr}(U\rho_a^{\rm eq})$,
  denotes the solvent reorganization energy, while its
variation satisfies
 $\la U^2 \ra - \la U \ra^2 \approx 2\lambda\kT$,
 the classical fluctuation-dissipation relation in
the high temperature limit.
  With a classical dielectric continuum model, Marcus had further related
 the solvent reorganization energy $\lambda$ to the
 static and high-frequency dielectric constants
  of the solvent.\cite{Mar56966,Mar64155,Mar85265}

The standard approach to ET rates is based on
correlation function formalism.\cite{Mar56966,Mar64155,Mar85265,%
Yan884842,Tan973485,Bix9935,%
Zus80295,Zus8329,Yan979361,Hyn85573,%
Gar854491,Fra85337,Wol871957,Spa873938,Spa883263,Spa884300}
 The second-order transfer correlation function
is related to the nonadiabatic rate
theory,\cite{Yan884842,Tan973485,Bix9935}
which in the static solvation limit assumes
the celebrated Marcus' rate expression,\cite{Mar56966,Mar64155,Mar85265}
 \be \label{mark0}
  k = \frac{V^2/\hbar}{\sqrt{\lambda\kT/\pi}}
      \exp\left[-\frac{(E^{\circ}+\lambda)^2}{4\lambda\kT}\right].
 \ee
Rates have also been formulated based on
fourth-order transfer correlation functions,
followed by certain resummation schemes
that partially account for the effect of
nonperturbative transfer coupling.\cite{Zus80295,Zus8329,Hyn85573,%
Gar854491,Fra85337,Wol871957,Spa873938,Spa883263,%
Spa884300,Yan884842,Yan979361}
The resulting rates do recover the celebrated
Kramers' turnover behavior,\cite{Kra40284,Han90251}
and are also able to relate the reaction adiabaticity to solvent
relaxation time scale.
However, these correlation function-based
 rate theories remain perturbative in nature.
To obtain rate expressions,
one shall also assume the separation
of time scales between the fast ET dynamics
and the slow solvation processes.\cite{Zus80295,Zus8329,Hyn85573,%
Gar854491,Fra85337,Wol871957,Spa873938,Spa883263,%
Spa884300,Yan884842,Yan979361} As results, the reaction Gibbs
free-energy $\Delta G^{\circ}$, obtained via the forward and
backward rates ratio $k/k'=\exp[-\Delta G^{\circ}/(\kT)]$, is
identical to the endothermicity parameter $E^{\circ}$ that
contains no explicit dependence on solvent environment.
 The parabolic dependence of $\log k$ on $E^{\circ}$,
 as shown by \Eq{mark0}, can be read as its
 dependence on $\Delta G^{\circ}$. This
 is the so-called Marcus' inversion
 behavior.\cite{Mar56966,Mar64155,Mar85265}

  Alternative approach to ET rates is via
 reduced density matrix,\cite{Fai80,Sto968126}
 defined formally as
   $\rho(t)\equiv {\rm tr}_{\rm B}\rho_{\rm T}(t)$,
 the trace of total density matrix
 over bath degrees of freedom.
  This is a quantum reduced EOM approach
 in which the transfer coupling is part of
 the system and can be treated exactly.
 However, the system-bath
 interaction, which in ET systems is
 neither weak nor Markovian,
 constitutes the major challenge in the general
 theory of quantum dissipation.\cite{Wei99,Yan05187,Xu05041103}

  It has been shown that an exact reduced dynamics
 theory, in terms of hierarchically coupled EOM,
 does exist in model Debye solvents that satisfies
 a semiclassical fluctuation-dissipation
 relation.\cite{Xu05041103,Tan89101} Based
 on this exact theory, we shall in this work construct an
analytical rate expression for the simple ET system, without
invoking such as resummation and timescale separation
approximations. As results, the present work will not just recover
for kinetic rates the celebrated  Kramers'
turnover\cite{Kra40284,Han90251} and Marcus'
inversion\cite{Mar56966,Mar64155,Mar85265}  behaviors, it will
also reveal for ET thermodynamics such as Gibbs free-energy and
entropy functions some interesting solvent dependent behaviors.

  The remainder of this paper is organized as follows.
 Section II treats an exact, nonperturbative
 theory of the reduced density matrix dynamics
 in Debye solvents. After a brief review of
 the hierarchical EOM formalism
 (\Sec{ththeoA}),\cite{Xu05041103,Tan89101}
 we construct a continued
 fraction Green's function theory of quantum dissipation
(\Sec{ththeoB}). We further utilize it, together with Dyson
equation technique, to evaluate analytically the reduced dynamics
of the simple ET system (\Sec{ththeoC}). Section \ref{ththeo2}
contributes to the development of reduced density matrix-based ET
rate theory. Numerical studies in \Sec{thnum} will demonstrate not
just for ET rates, but also for ET reaction (equilibrium)
thermodynamics, their dependence on solvent environment. Finally,
\Sec{thsum} concludes the paper.

\section{Exact reduced dynamics in Debye solvents}
\label{ththeo}

\subsection{Hierarchical equations of motion formalism}
\label{ththeoA}

   To describe the hierarchical EOM
 for reduced density matrix,\cite{Xu05041103,Tan89101}
 let us recast the total ET Hamiltonian (\Eq{HT})
 in the stochastic bath interaction
 picture,
 \be\label{HTt}
  H_{\rm T}(t) = H + H'(t),
 \ee
 with $H$ and $H'(t)$ representing
 the reduced system Hamiltonian
 and the stochastic system-bath coupling,
 respectively.
 \be \label{Hsys}
  H = (E^{\circ}+\lambda)|b\ra\la b|  + V(|a\ra\la b|+|b\ra\la a|),
 \ee
 \be \label{Hsb}
  H'(t) = [U(t)-\lambda] |b\ra\la b|.
 \ee
 The stochastic solvation coordinate, \be \label{solvU}
 U(t)\equiv e^{ih_at/\hbar}Ue^{-ih_at/\hbar}
   e^{ih_at/\hbar}(h_b-h_a)e^{-ih_at/\hbar} ,
\ee
is assumed to be of Gaussian statistics.
Thus, the effects of solvent on the ET system are
completely determined by the solvent reorganization energy,
 \be \label{lamb}
   \lambda = \la U(t) \ra \equiv
       {\rm tr} [U(t)\rho_a^{\rm eq} ] = \la U \ra,
 \ee
 and the solvation correlation function,
 \be \label{Ct}
    C(t-\tau) = \la [U(t)-\lambda][U(\tau)-\lambda] \ra.
 \ee

  In this work, we focus on the ET system in a Debye solvent
 (also called the Drude model), characterized by the following
 form of solvation response function,\cite{Wei99}
 \be \label{resp}
  i\la [U(t),U(0)] \ra = -2\,{\rm Im}\,C(t)
   = 2\Theta(t)\hbar\lambda\gamma e^{-\gamma t}.
 \ee
 Here, $\Theta(t)$ is the Heaviside step function
 and $\gamma^{-1} \equiv \tau_{\rm L}
 = \tau_{\rm D}(\varepsilon_{\infty}/\varepsilon_0)$,
  with $\tau_{\rm D}$ being the Debye time parameter,
  and $\varepsilon_0$ ($\varepsilon_{\infty}$)
  the static (high-frequency) dielectric
constant of the solvent. The corresponding solvation correlation
function in the semiclassical high-temperature limit
reads\cite{Wei99} \be \label{debyeC}
   C(t)  \approx \lambda(2\kT - i\hbar\gamma)e^{-\gamma t} .
\ee
 For this model, the exact reduced density
 matrix dynamics has been constructed,
 in terms of\cite{Xu05041103,Tan89101}
 \be\label{dotrhon}
  \dot\rho_n =-(i\L+n\gamma) \rho_n
    -i\B\rho_{n+1}-in\A\rho_{n-1}; \ \ n\geq 0,
 \ee
 which hierarchically couple the
 $\rho\equiv \rho_0$ of primary interest
 and a set of auxiliary system operators
 $\{\rho_n; n=1,2,\cdots\}$.
  The initial conditions are
 $\rho_{n}(0)=\rho(0)\delta_{n0}$, and
 \bsube \label{LAB}
 \bea
  \L \hat O &\equiv& \hbar^{-1} [H,\hat O],
 \label{calL} \\
  \A\hat O &\equiv&
 \frac{2\lambda\kT}{\hbar^2}[|b\ra\la b|, \hat O]
  - i\frac{\lambda\gamma}{\hbar} \{|b\ra\la b|,\hat O\} ,
 \label{calA} \\
  \B\hat O &\equiv& [|b\ra\la b|, \hat O].
 \label{calB}
 \eea
 \esube
 Here, $\{\cdot,\cdot\}$ denotes anticommutator.
 It has been shown\cite{Xu05041103} that
 the individual auxiliary operator $\rho_{n>0}$
 accounts for the $2n^{\rm th}$-order
 system-bath interaction contribution to the
 reduced dynamics of the primary interested $\rho$;
 see also the comments in the last paragraph of
 \Sec{ththeoB}.

\subsection{Continued fraction Green's function formalism}
 \label{ththeoB}
 Introduce the propagators $\{\U_n(t);n=0,1,\cdots\}$:
 \be \label{Un}
  \rho_n(t)\equiv e^{-n\gamma t}\U_n(t)\rho(0);
 \ \  {\rm with\ \ } \U_n(0)=\delta_{n0}.
 \ee
 Equations \ref{dotrhon} read now
\[
  \dot\U_n(t)=-i\L \U_n(t)
  -i\B e^{-\gamma t}\U_{n+1}(t)
   -in\A e^{\gamma t}\U_{n-1}(t),
\]
which in the Laplace-domain are
 \bea \label{Uns}
 (s\!+\!i\L)\ti\U_n(s)\!+\!%
 i\B\ti\U_{n\!+\!1}(s\!+\!\gamma)%
  \!+\!in\A\ti\U_{n\!-\!1}(s\!-\!\gamma)%
 \!=\!\delta_{n0}.
\eea
Define the Green's functions $\{\G^{(n)}(s); n\geq 0\}$ via
\bsube \label{Gndef}
\bea
  \ti\U_0(s) &\equiv& \G^{(0)}(s)  \equiv \G(s),
\label{Grn0} \\
  \ti\U_n(s) &\equiv& -in\G^{(n)}(s)\A
  \ti\U_{n-1}(s-\gamma); \ \ n>0.
\label{Gn}
\eea
\esube
 These equations will lead to
\bsube \label{GPi}
\be \label{Gs}
  \G^{(n)}(s) = \frac{1}{s+i\L+\Pi^{(n)}(s)};
\ \ \  n\geq 0,
\ee
 with
 \be \label{Pis}
 \Pi^{(n)}(s) \equiv (n+1)\B\G^{(n+1)}(s+\gamma)\A.
 \ee
 \esube
 The above equations, which can be recast as
 \be \label{Pns}
  \Pi^{(n)}(s) = \B
   \frac{n+1}{s+\gamma+i\L+\Pi^{(n+1)}(s+\gamma)}\A,
 \ee
 constitute the infinite continued fraction formalism
 for evaluating each individual $\Pi^{(n)}(s)$ or  $\G^{(n)}(s)$.

  The Green's function $\G^{(0)}(s)\equiv \G(s)$
and its associated $\Pi^{(0)}(s)\equiv \Pi(s)$
are of the primary interest.
The former resolves the reduced density matrix evolution
(cf.~\Eq{Grn0} and \Eq{Un} with $n=0$),
\be \label{rhos0}
  \ti\rho(s) \equiv \int_0^{\infty}\!\!dt\, e^{-st}\rho(t)
  = \G(s)\rho(0).
\ee
This equation can be react as (cf. \Eq{Gs} at $n=0$)
\be \label{rhos}
    s\ti \rho(s) - \rho(0) =
   -i\L \ti\rho(s)-\Pi(s)\ti\rho(s),
\ee
which in the time-domain reads
\be \label{rhot}
  \dot\rho(t) = -i\L\rho(t) - \int_0^{t}\!d\tau \,
     \hat\Pi(t-\tau)\rho(\tau).
\ee
Therefore,
\be \label{Ptst}
  \Pi(s)=\Pi^{(0)}(s) = \int_0^{\infty}\!dt\, e^{-st}\hat\Pi(t),
\ee
represents the memory kernel in the Laplace domain.

  The initial input for the inverse recursive
 evaluation of $\Pi(s)$ (\Eq{Pns})
 can be chosen based on the
 following observation. Each $\A$ is of second order in the
system-bath coupling; thus the leading contribution of $\Pi^{(n)}$
to the required $\Pi$ is of the $(2n)^{\rm th}$ order. Moreover,
as the mathematical nature of continued fraction, convergency is
also guaranteed practically for arbitrary strength and timescale
of system-bath coupling. We can therefore set $\Pi^{(N+1)}=0$,
with a sufficiently large $N$, to initiate the
 inverse recursive procedure, and
 evaluate $\Pi^{(n)}(s+n\gamma)$; first at $n=N$,
 then $N-1$, and so on, until the
 required $\Pi^{(0)}(s)=\Pi(s)$ is reached.

\subsection{Evaluation of tensor elements}
\label{ththeoC}
  The tensor element of an superoperator (or Liouville-space operator)
${\cal O}$ is defined in the double-bracket notation as\cite{Fan5774,Muk95}
\be
 {\cal O}_{jj',kk'}\equiv \la\la jj'|{\cal O}|kk'\ra\ra ,
\ee
so that
\be
 {\cal O} = \sum_{jj',kk'} {\cal O}_{jj',kk'}|jj'\dd kk'|.
\ee
For a two-level system considered in this work, each tensor
has $2^4=16$ elements.  That ${\cal O}$ is Hermite implies
${\cal O}_{jj',kk'}={\cal O}^{\ast}_{j'j,k'k}$.
Apparently, all $\Pi^{(n)}$ and $\G^{(n)}$  are Hermite.

  To analyze the tensor elements of $\Pi^{(n)}$ (\Eq{Pis}), let us
first examine $\A$ and $\B$,
defined by \Eqs{calA} and \ref{calB}, respectively.
They are found to be diagonal, with the nonzero elements
of $\A_{ba,ba}=-\A^{\ast}_{ab,ab}=
     \lambda(2\kT -i\hbar\gamma)/\hbar^2$,
$\A_{bb,bb}=-i2\lambda\gamma/\hbar$,
and $\B_{ba,ba}=-\B^{\ast}_{ab,ab}=1$.
As results, the only nonzero elements in $\Pi^{(n)}$ (\Eq{Pis}) are
\be \label{xyz}
   x^{(n)} \equiv  \Pi^{(n)}_{ba,ba}, \
   y^{(n)} \equiv  \Pi^{(n)}_{ba,ab}, \
   z^{(n)} \equiv  \Pi^{(n)}_{ba,bb},
\ee
and their Hermitian conjugate elements,
and they are related to the Green's function tensor elements,
\be \label{XYZ}
   X^{(n)} \equiv  \G^{(n)}_{ba,ba}, \
   Y^{(n)} \equiv  \G^{(n)}_{ba,ab}, \
   Z^{(n)} \equiv  \G^{(n)}_{ba,bb} ,
\ee
by [denoting $\eta\equiv \lambda(2\kT-i\hbar\gamma)/\hbar^2$]
\bsube \label{finaldebye}
\bea
   x^{(n)}(s) &=& \eta (n+1) X^{(n+1)}(s+\gamma),
    \ \  \ \
\\
  y^{(n)}(s) &=& -\eta^{\ast}(n+1) Y^{(n+1)}(s+\gamma),
\\
  z^{(n)}(s) &=& (\eta-\eta^{\ast})(n+1) Z^{(n+1)}(s+\gamma).
\eea
\esube

 To evaluate the involving Green's function elements via \Eq{Gs},
we apply the Dyson equation,
 \be \label{dyson}
  \G=\bar\G - \bar\G(i\L'+\Pi')\G ,
 \ee
 with $\bar\G$ being the diagonal contribution, and $\L'$ and
$\Pi'$ the off-diagonal parts of the involving $\L$ and $\Pi$ in
\Eq{Gs}, respectively. Here and in \Eqs{finalZ} and \ref{allaug}
follows, the common superscript $(n)$ and argument $s$ in both
sides of equations are implied. After some elementary algebra, we
obtain
 \bsube \label{finalXYZ}
  \bea \label{finalX}
  X &=& \frac{ \alpha^{\ast}+\beta^{\ast} }
             {|\alpha + \beta|^2 - |\beta - y|^2},
 \\ \label{finalY}
 Y &=& \frac{\beta-y}{|\alpha + \beta|^2 - |\beta - y|^2},
\\ \label{finalZ}
 Z &=& -\frac{1}{s}
   \bigl[(z-iV/\hbar) X + (z^{\ast}+iV/\hbar) Y\bigr],
\eea
 \esube
with
 \bsube \label{allaug}
 \bea
   \alpha &\equiv&  s+(i/\hbar)(E^{\circ}+\lambda)+x,
 \label{alpha} \\
    \beta &\equiv& s^{-1}(V/\hbar)^2(2+i\hbar z/V).
 \label{beta}
 \eea
 \esube
 We have thus established from \Eq{Gs} the expressions of
$\{X,Y,Z\}^{(n)}(s)$ in terms of $\{x,y,z\}^{(n)}(s)$,
 which together with \Eqs{finaldebye},
 constitute an analytical and exact
 formalism for the inverse recursive evaluation of the reduced
 dynamics in Debye solvents. In the following section, we shall
show that the ET reaction rate can be expressed in terms of
$\{x,y,z\}$, i.e., the nonzero elements of dissipative memory
kernel $\Pi$ in the Laplace domain.

\section{Electron transfer rate: Reduced-density-matrix formalism}
\label{ththeo2}
 We are now in the position to construct
the reduced density matrix approach to ET rates. Let us start with
$\ti\rho(s)$ [\Eq{rhos}], where $\rho(t=0)=|aa\ra\ra$.
 By separating $\ti\rho(s)$ into population
 vector $\ti{\mbox{\boldmath $P$}}
  = [\ti\rho_{aa},\ti \rho_{bb}]^{\rm T}$
 and coherent vector $[\ti\rho_{ab},\ti \rho_{ba}]^{\rm T}$ components,
and then using \Eq{rhos} to eliminate the latter,
 we obtain the ET kinetic equations in Laplace domain as
 \be
 \label{ratePs}
 s\ti{\mbox{\boldmath $P$}}(s)-{\mbox{\boldmath $P$}}(0) =
   K(s)\ti{\mbox{\boldmath $P$}}(s),
 \ee
with
\be\label{rateKs}
   K(s) =  T_{\mbox{\tiny PC}}(s+T_{\mbox{\tiny CC}})^{-1}
     T_{\mbox{\tiny CP}} - T_{\mbox{\tiny PP}} .
\ee
Here, $T_{\mbox{\tiny PC}}$, $T_{\mbox{\tiny CC}}$,
$T_{\mbox{\tiny CP}}$, and $T_{\mbox{\tiny PP}}$
denote the coherence-to-population, coherence-to-coherence,
population-to-coherence, and population-to-population
transfer matrices involved in \Eq{rhos},
respectively.
Tensor analysis (cf.\ \Sec{ththeoC}) results in
$T_{\mbox{\tiny PP}}=0$, and
\bsube \label{allT}
\be \label{Tpc}
 T_{\mbox{\tiny PC}} =
    iV\left[\begin{array}{cc} -1 & 1 \\ 1 & -1 \end{array}\right],
\ \ 
 T_{\mbox{\tiny CP}} = T_{\mbox{\tiny PC}} +
     \left[\begin{array}{cc} 0 & \Pi^{\ast}_{10,11} \\ 0 & \Pi_{10,11}
  \end{array}\right],
\ee
\be \label{Tcc}
  T_{\mbox{\tiny CC}}
 =
    i(E^{\circ}+\lambda)
    \left[\begin{array}{cc} -1 & 0 \\ 0 & 1 \end{array} \right]
   + \left[\begin{array}{cc}
            \Pi^{\ast}_{10,10} & \Pi^{\ast}_{10,01} \\
             \Pi_{10,01} & \Pi_{10,10}
   \end{array} \right] .
\ee
\esube

 Note that \Eq{ratePs} in time domain  reads
\be \label{ratePt0}
  \dot{\mbox{\boldmath $P$}}(t) = \int_{0}^{t}\!d\tau \hat K(t-\tau)
    {\mbox{\boldmath $P$}}(\tau).
\ee Thus $K(s)$ is the resolution or the Laplace-transform of the
memory rate kernel $\hat K(t)$. The total population conservation
implies the relation $K_{aj}+K_{bj}=0$;
 thus,  \Eq{ratePt0} is equivalent to
 \be \label{ratePt}
   \dot P_a(t)
   = - \int_{0}^{t}\!d\tau \hat k(t-\tau)P_a(\tau)
     + \int_{0}^{t}\!d\tau \hat k'(t-\tau)P_b(\tau) .
 \ee
The forward and backward rate resolutions are 
\bsube \label{kdef}
\bea
  k(s)&=& -K_{aa}(s) = \int_0^{\infty}\!dt\, e^{-st} \hat k(t),
\\
  k'(s)&=& K_{ab}(s) = \int_0^{\infty}\!dt\, e^{-st}\hat k'(t).
\eea
\esube
Together with \Eq{rateKs}, \Eqs{allT} and $T_{\mbox{\tiny PP}}=0$,
we obtain
\bsube\label{finalK}
\be \label{Ks}
  k(s) = \frac{2|V|^2}{\hbar^2 }{\rm Re}
    \frac{\alpha(s) + y(s)}{|\alpha(s)|^2 - |y(s)|^2},
 \ee
  and
 \be \label{Kbs}
  k'(s) = \frac{2|V|^2}{\hbar^2}
    {\rm Re} \frac{[\alpha(s) + y(s)][1-i\hbar z^{\ast}(s)/V]}
   {|\alpha(s)|^2 - |y(s)|^2}.
 \ee
\esube
Here $\alpha(s) = s+(i/\hbar)(E^{\circ}+\lambda)+x(s)$
is the same as \Eq{alpha}, while
$x\equiv \Pi_{ba,ba}\equiv x^{(0)}$,
$y\equiv \Pi_{ba,ab}\equiv y^{(0)}$,
and $z\equiv \Pi_{ba,bb}\equiv z^{(0)}$
are the same as \Eqs{xyz} at $n=0$.

 The above formulations relate
the ET reaction rate resolutions $k(s)$ and $k'(s)$, to
 the memory dissipation kernel resolution $\Pi(s)$;
cf.\ \Eqs{rhos} and \ref{rhot}. The analytical expressions for the
 inverse recursive evaluation of the $\Pi(s)$
 tensor elements had been
 detailed in \Sec{ththeoB} and \Sec{ththeoC}. We have thus
 completed the analytical expressions for
 the frequency-dependent rates of
 ET in Debye solvents.

\section{Rates versus thermodynamics: Numerical demonstrations}
\label{thnum}
  For numerical study, we focus on the
ET rate constants, $k\equiv k(s=0)$ and $k'\equiv k'(s=0)$, which
amount to the integrated memory rate kernel (cf.\ \Eq{kdef}).
 It is easy to show that in the weak
 transfer coupling (small $V$) and
 slow solvation (small $\gamma$) limit,
 the present rate constant $k$
 (\Eq{Ks} at $s=0$) recovers the celebrated Marcus' ET rate
  expression of \Eq{mark0}.
 We shall also be interested in the reaction
 Gibbs free-energy, $\Delta G^{\circ} = -\kT \ln (k/k')$,
 entropy $\Delta S^{\circ}=-\partial \Delta G^{\circ}/\partial T$,
 and enthalpy
  $\Delta H^{\circ}= \Delta G^{\circ}+T\Delta S^{\circ}$.
  We shall demonstrate how the rate constants
 and reaction thermodynamics functions depend on the model
 parameters, reaction endothermicity $E^{\circ}$, solvent
reorganization energy $\lambda$, and longitudinal relation time
$\tau_{\rm L}=1/\gamma$. The other two parameters of the model are
set to be $T=298$ K and $V=1$ kJ/mol, unless being further
specified. In fact, the value of $V$ can be considered as the unit
that scales other parameters used. The temperature should also
varies around $T=298$ K in order to numerically evaluation of
entropy.

Note that in optical spectroscopy, one often uses a dimensionless
 parameter, $\kappa^{-1} = \tau_{\rm L}\sqrt{2\kT\lambda}/\hbar$, to
 measure the slow ($\kappa\ll 1$) and the fast ($\kappa \gg 1$)
 solvent modulation limit.
  For $\lambda= 3$ kJ/mol at $T=298$ K,
 that $\kappa=1$ corresponds to $\tau_{\rm L} = 16.5$ fs,
 while the typical ET solvation correlation timescale is
 of picoseconds. We will show that the parameter $\kappa$
 can also be used in ET rate problem.

   Figure \ref{fig1} depicts ET rate $k$ as the function of solvent
 relaxation time $\tau_{\rm L}=1/\gamma$
 at four representing values of endothermicity,
 $E^{\circ} = 0, -1, -3,$ and $-5$ kJ/mol.
 The solvent reorganization energy $\lambda=3$ kJ/mol.
 In general, the ET rate consists of the barrier crossing and
 the coherent tunneling contributions.
 When $E^{\circ} = 0$ (symmetric case),
 the system is in Fermi resonance and the ET is dominant by
 coherent tunneling.
 The observed rate in this case exhibits the
 motional narrowing behaviors:\cite{Kub69101,Yan884842}
 the faster the solvent modulation is,
 the larger the coherent resonant tunneling rate will be.
 In a nonsymmetric ($E^{\circ} \neq 0$) system, the barrier crossing
 is significant and the observed ET rate exhibits clearly
 the Kramers' turnover behaviors.\cite{Kra40284,Han90251}

    The rates observed in the slow solvent modulation
 (large $\tau_{\rm L}$ which amounts to large viscosity)
 region of \Fig{fig1}
 are closely related to the Marcus' inversion behaviors.
 Figure \ref{fig2} depicts the ET rate, in term of
 $\ln (k/k_{\rm max})$, as the function of reaction
 endothermicity $E^{\circ}$, in the slow modulation
 region ($\tau_{\rm L}=10$ ps), at two specified values of
 transfer coupling strength, $V=0.01$ and 1 kJ/mol.
  When the transfer coupling is small ($V=0.01$ kJ/mol),
  the rate in the slow solvent modulation regime
  does have the Marcus' nonadiabatic form,
 $\ln (k/k_{\rm max}) = -(E^{\circ}+\lambda)^2/(4\kT\lambda)$
 of \Eq{mark0};
 cf.\ the thin-solid vs.\ thin-dash curves.
 In the case of $V=1$ kJ/mol, however, the rate deviates significantly
 from the Marcus' expression. The poorly-fitted parabolic function there,
 $-(E^{\circ}- E^{\circ}_{\rm max})^2/(\zeta 4\kT\lambda)$
 (dash-curve), is found to be of
 $E^{\circ}_{\rm max} =-2.4$ kJ/mol and $\zeta=0.3$.

   Figure \ref{fig3} shows the 3d-plot of reaction Gibbs free-energy
 $\Delta G^{\circ}$  as the function of
 ($\lambda,\tau_{\rm L})$, exemplified
 at $T=298$ K with $E^{\circ}=-3$ kJ/mol and $V=1$ kJ/mol.
  Reported in \Fig{fig4} are some representing 2d-slices of \Fig{fig3}
  for $\Delta G^{\circ}$, together
  with the numerically evaluated reaction
  entropy $\Delta S^{\circ}$ and enthalpy $\Delta H^{\circ}$.
    The observed features here are listed as follows.

(i) The basic symmetry requirements,
  such as $\Delta G^{\circ}(-E^{\circ})=-\Delta G^{\circ}(E^{\circ})$
    which implies also $\Delta G^{\circ}(E^{\circ}=0)=0$,
   hold in general;

(ii)
 $\Delta G^{\circ}$, $\Delta H^{\circ}$ and $E^{\circ}$ are of same sign
   and $|\Delta H^{\circ}| \geq |\Delta G^{\circ}| \geq |E^{\circ}|$,
   implying that the enthalpy and the entropy play opposite roles
   on the reaction Gibbs free-energy;

(iii)
 The dependence of $\Delta G^{\circ}$ on ($\lambda,\tau_{\rm L}$),
    as shown in \Fig{fig3}, is qualitatively similar to
    that in the weak transfer coupling regime;

(iv)
  In general, $\Delta G^{\circ}$ approaches to a constant
   in both the fast ($\kappa \gg 1$)
   and slow ($\kappa \ll 1$) modulation regimes.
   The smaller $V$ is, the closer $\Delta G^{\circ}$ to $E^{\circ}$ will
   be.
  For the system demonstrated in \Fig{fig3} (or \Fig{fig4})
   where $E^{\circ}=-3$ kJ/mol with $V=1$ kJ/mol,
  $|\Delta G^{\circ}(\kappa\gg 1)|
  > |\Delta G^{\circ}(\kappa\ll 1)|$;
  however the sign could be opposite for
  small $|E^{\circ}|$ systems (not shown here);

(v)
  In the modest modulation regime ($\kappa \sim 1$),
   the reaction thermodynamic functions
  ($\Delta G^{\circ}$,  $\Delta H^{\circ}$,
   and $\Delta S^{\circ}$) exhibit certain nonlinear
   dependence on ($\lambda,\tau_{\rm L}$).
   In particular, the magnitudes of these ET reaction
   thermodynamic functions are of maximum values
   around $\kappa \sim 1$ when $\lambda > |E^{\circ}|$.

   Interestingly, the observed Kramers' turnover
   of rates occurs also around the $\kappa \sim 1$
   (intermediate friction) region. The relevant
  forward and backward rate constants are given in
 \Fig{fig5}.
  This may at least partially account
  for the  nonmonotonic
  dependence of thermodynamics functions
  on the solvent environment in this region,
  as they are related to the rates via the detailed balance relation.

\section{Summary}
\label{thsum}
  In summary, we have constructed a formally exact,
nonperturbative ET rate theory in terms of the resolution of the
memory dissipation kernel (cf.\ \Eqs{finalK} with \Eqs{Ptst} and
\ref{xyz}). For Debye solvents in which the solvation correlation
is characterized by an exponential function (\Eq{debyeC}), the ET
rates, or rate resolutions in general (\Eqs{finalK}), can be
evaluated readily via the
 inverse-recursion formalism (cf.\
\Eqs{finaldebye} with \Eqs{finalXYZ}). Not only does it recover
the celebrated Marcus' inversion and Kramers' turnover behaviors
of the ET rates, the present formalism also provides a microscopic
theory for the effects of solvent environment on the ET
thermodynamics.  The dependence of reaction thermodynamics on
solvent environment is found to be quite dramatic, especially in
the region where the  Kramers' turnover occur(s). This
observation suggests the possibility of utilizing the
thermodynamics data to extract such as the solvation correlation
time parameter (cf.\ the left panel of \Fig{fig4}).

 This work has also developed a
nonperturbative theory of reduced density matrix dynamics, in
terms of continued fraction Green's function (\Sec{ththeoB}). This
formalism can be used together with Dyson equation technique for
efficient and analytical evaluation of reduced dynamics
(\Sec{ththeoC}). The present formalism is exact for the Debye
solvents (\Eq{debyeC}). However, as the semiclassical
fluctuation-dissipation theorem is involved
 in \Eq{debyeC}, the
reduced density matrix and rates may become negative if $\kT \ll
(|V|^2+\frac{1}{4}|E^{\circ}|^2)^{1/2}$.
Generalization of the present continued fraction Green's function
approach to non-Debye solvents at arbitrary temperature
is feasible and will be developed in future.

\begin{acknowledgments}
 Support from the RGC Hong Kong and the NNSF of
  China (No.\ 50121202, No.\ 20403016 and No.\ 20533060)
 is acknowledged.
\end{acknowledgments}


\clearpage
 \begin{figure}
 \caption{
   Electron transfer rates  as the functions of solvent longitudinal
   relaxation time $\tau_{\rm L} \equiv 1/\gamma$, for some
   specified values of endothermicity $E^{\circ}$. The solvent
   reorganization energy, transfer coupling
   strength, and temperature are $\lambda=3$ kJ/mol, $V=1$ kJ/mol,
   and $T=298$ K,  respectively.
 }
 \label{fig1}
 \end{figure}

 \begin{figure}
 \caption{The normalized rates, in
   the slow modulation regime ($\tau_{\rm L}=10$ ps),
   as the functions of $E^{\circ}$.
   The Marcus' parabolic relation, $\ln (k/k_{\rm max}) =
   -(E^{\circ}+\lambda)^2/(4\kT\lambda)$,
   where $\lambda=3$ kJ/mol, is recovered in the case of $V=0.01$
   kJ/mol, but not for $V=1$ kJ/mol. The latter case  neither fits
   well with a parabolic function (dash curve); see the
   text for details.
 }
 \label{fig2}
 \end{figure}

 \begin{figure}
 \caption{
   Reaction Gibbs free-energy $\Delta G^{\circ}$
 as the function of solvent parameters
    $(\lambda,\tau_{\rm L})$, for an ET system with $E^{\circ}=-3$
    kJ/mol and $V=1$ kJ/mol at $T=298$ K.
 }
 \label{fig3}
 \end{figure}

 \begin{figure}
 \caption{
   Reaction thermodynamics functions $\Delta G^{\circ}$,
   $\Delta S^{\circ}$, and $\Delta H^{\circ}$:
   (a) as the functions of $\lambda$ at some selected
      values of $\tau_{\rm L}$;
   (b) as the functions of $\tau_{\rm L}$ at some selected values
       of $\lambda$. The ET system is same as that of \Fig{fig3}.
 }
 \label{fig4}
 \end{figure}

 \begin{figure}
 \caption{
    The forward and backward rate constants,
     $k$ (upper panels) and $k'$ (lower panels),
   relevant to the reaction thermodynamics functions
   in \Fig{fig4}.
 }
 \label{fig5}
 \end{figure}

\clearpage
\begin{center}
\centerline{\includegraphics[width= 0.7\columnwidth,angle=-90]
 {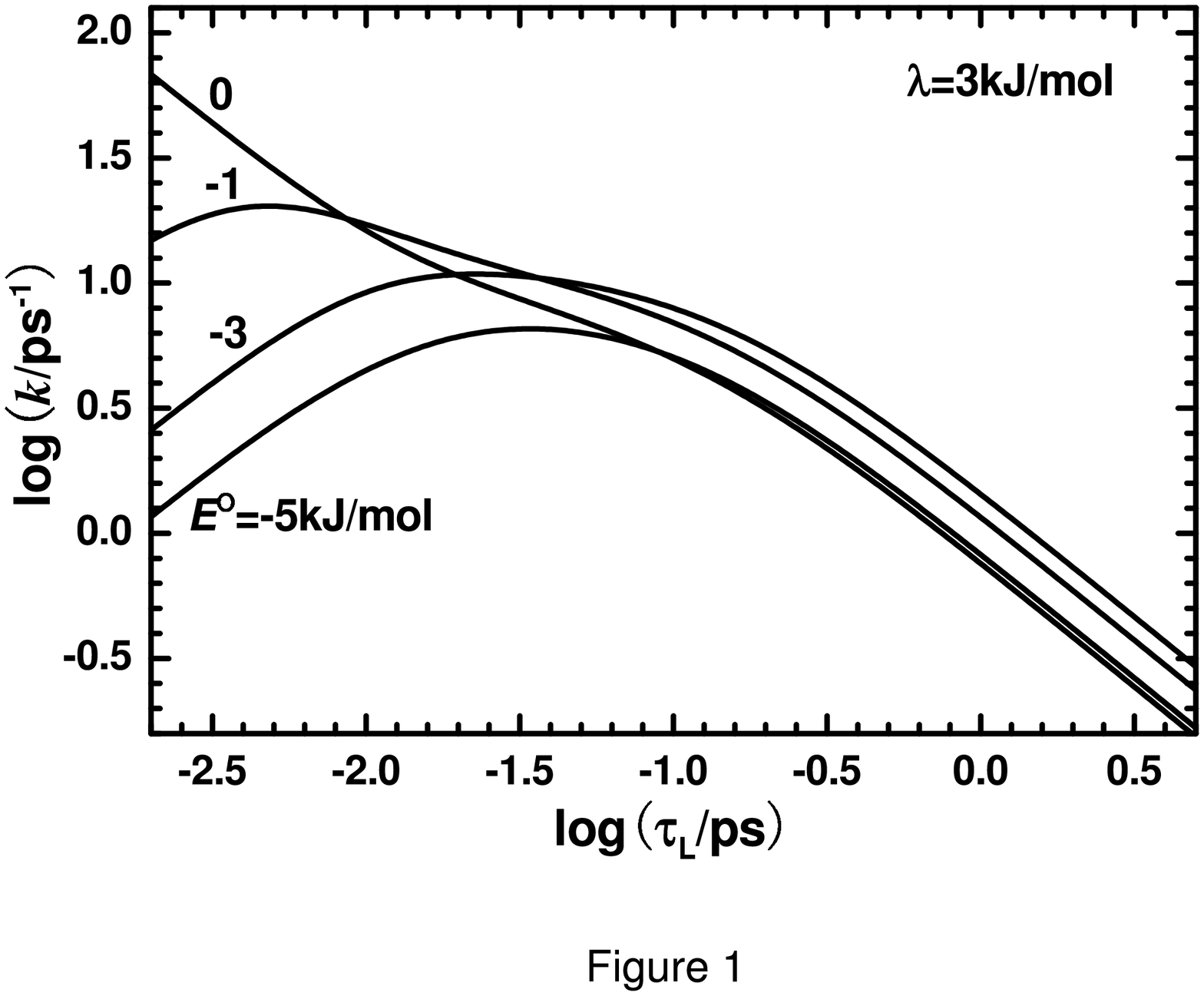}}
\centerline{\includegraphics[width= 0.7\columnwidth,angle=-90]
 {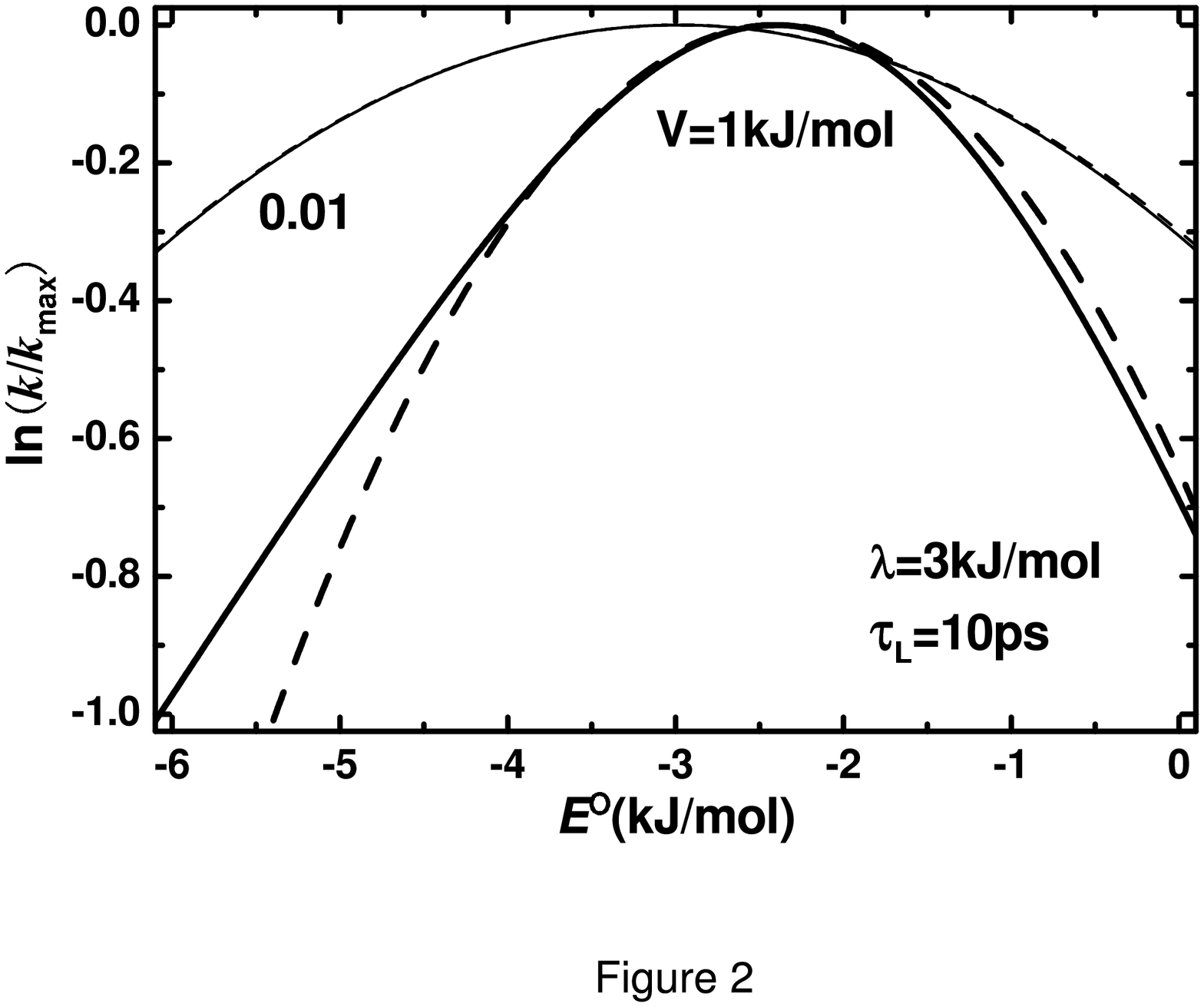}}
\centerline{\includegraphics[width= 0.7\columnwidth,angle=-90]
 {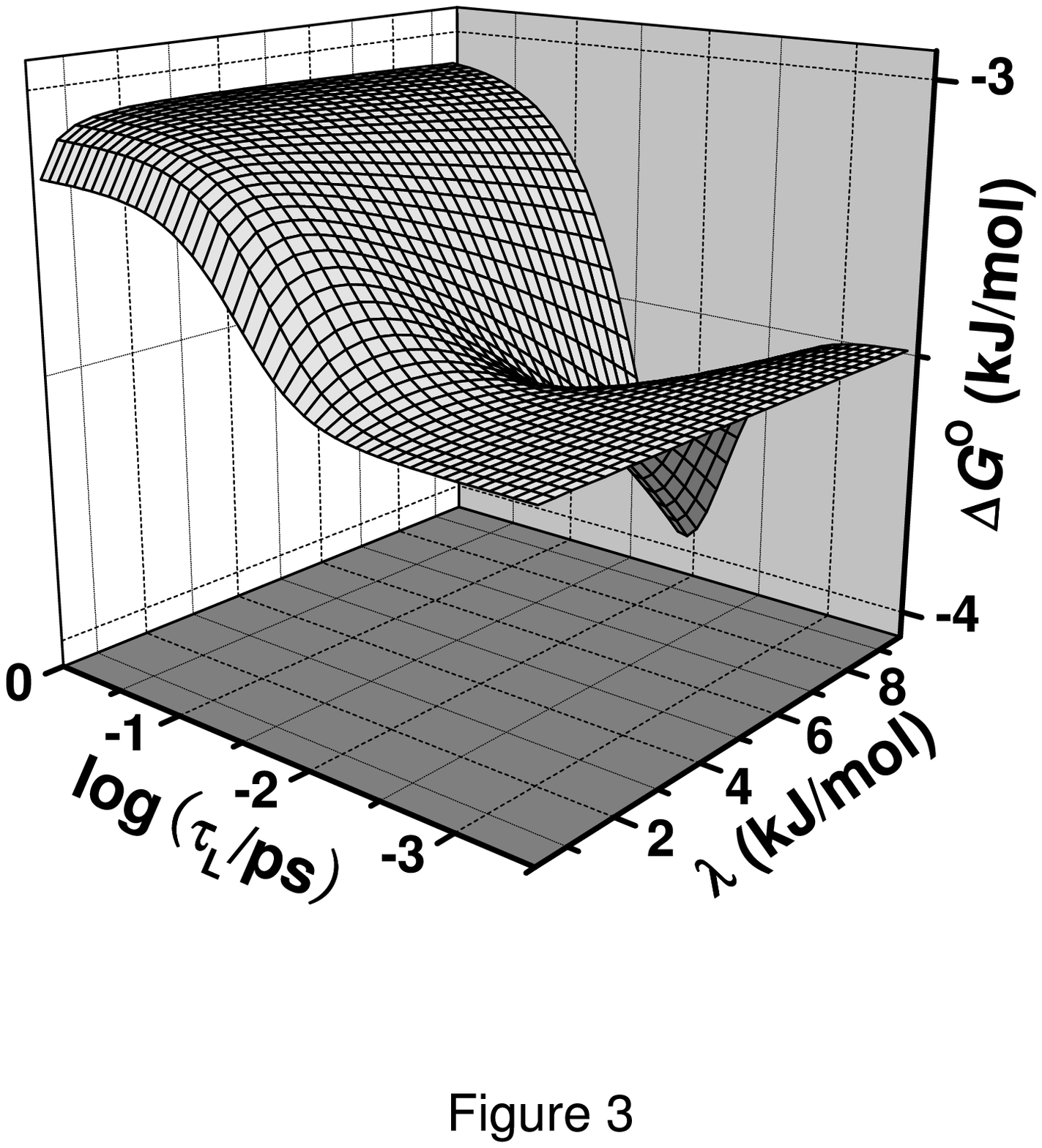}}
\centerline{\includegraphics[width= 0.7\columnwidth,angle=-90]
 {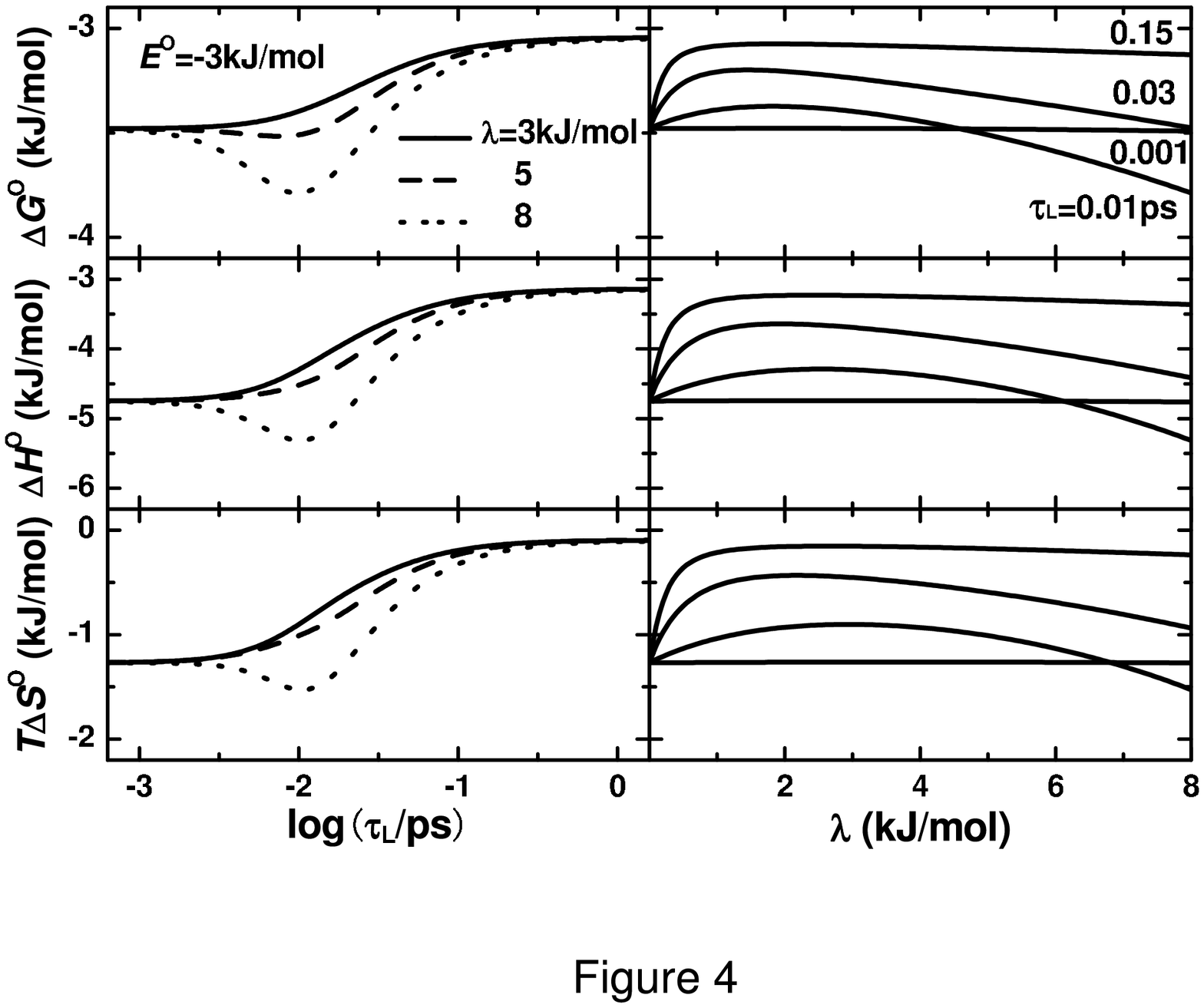}}
\centerline{\includegraphics[width= 0.7\columnwidth,angle=-90]
 {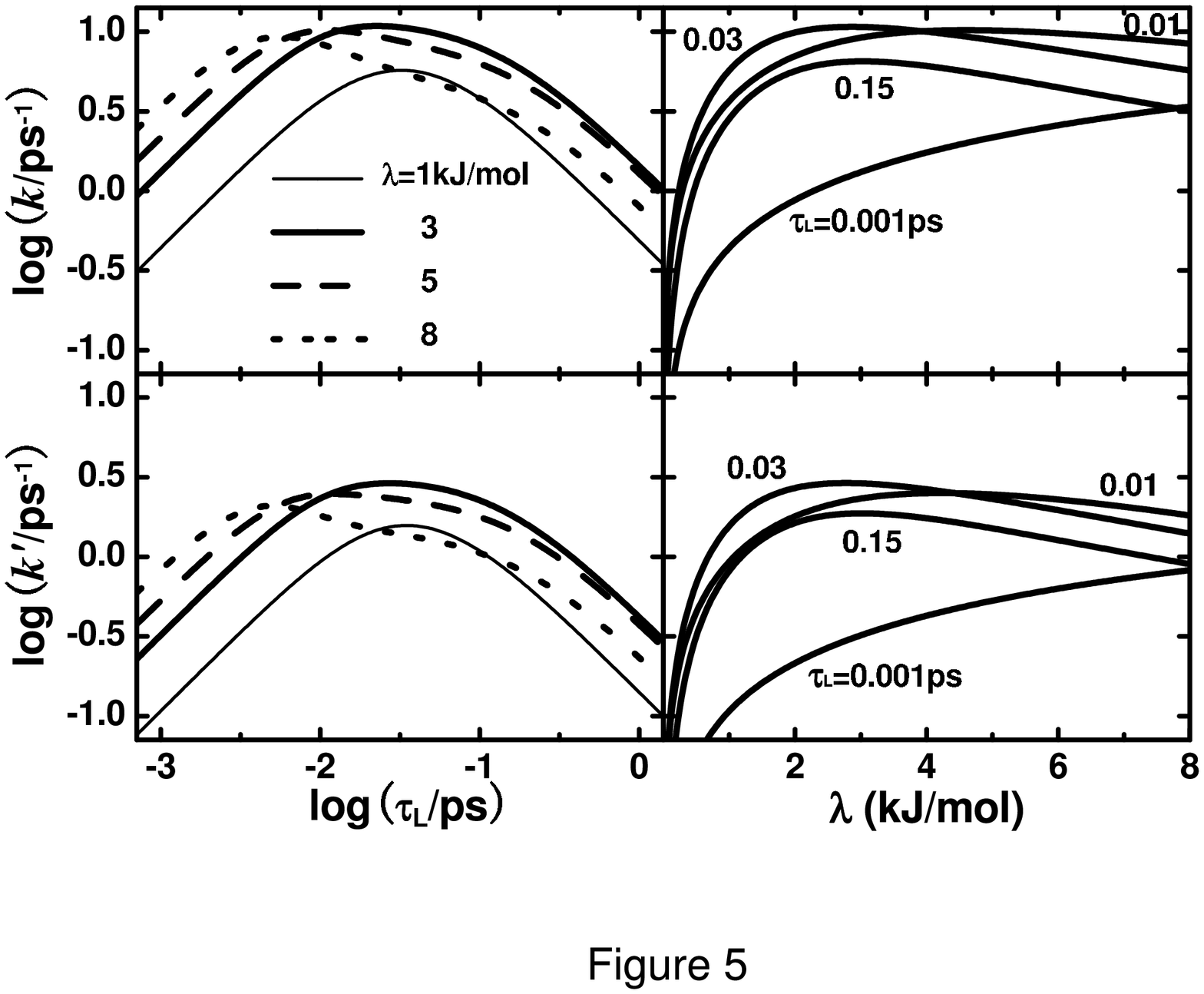}}
\end{center}

\end{document}